\documentclass[twocolumn,amsfonts,showpacs,superscriptaddress,nofootinbib]{revtex4-1}
\usepackage{amsmath}
\usepackage[dvipdfmx]{graphicx}
\usepackage{amssymb}
\usepackage{bm}
\usepackage{color}
\usepackage{color}
\usepackage{comment} 
\usepackage{ulem} 

\newcommand{\bc}{\begin{center}}
\newcommand{\ec}{\end{center}}
\def\ba#1{\begin{array}{#1}\displaystyle}
\newcommand{\ea}{\end{array}}

\newcommand{\beq}{\begin{equation}}
\newcommand{\eeq}{\end{equation}}
\newcommand{\beqa}{\begin{eqnarray}}
\newcommand{\eeqa}{\end{eqnarray}}
\newcommand{\beqs}{\begin{eqnarray*}}
\newcommand{\eeqs}{\end{eqnarray*}}
\newcommand{\no}{\nonumber}
\newcommand{\n}{\nonumber\\}
\newcommand{\bi}{\begin{itemize}}
\newcommand{\ei}{\end{itemize}}

\def\lt#1{\left#1}
\def\rt#1{\right#1}

\def\h#1{\hat{#1}}
\def\b#1{\bar{#1}}
\def\frc#1#2{\frac{#1}{#2}}
\newcommand{\p}{\partial}

\newcommand{\bra}{\langle}
\newcommand{\ket}{\rangle}

\newcommand{\R}{{\mathbb{R}}}

\newcommand{\Or}{{\cal O}}

\newcommand{\ep}{\epsilon}

\newcommand{\dd}{{\rm d}}
\newcommand{\ii}{{\rm i}}

\newcommand{\prob}{\mathbb{P}}
\newcommand{\expec}{\mathbb{E}}

\DeclareMathOperator{\ad}{ad}
\DeclareMathOperator{\Ad}{Ad}

\def\eqref#1{(\ref{#1})}

\begin{document}

\title{Diffusion and signatures of localization in stochastic conformal field theory}

\author{Denis Bernard}
\affiliation
{Laboratoire de Physique Th\'eorique de l'Ecole Normale Sup\'erieure de Paris,\\ CNRS, ENS \& PSL Research University, UPMC \& Sorbonne Universit\'es, France.
}

\author{Benjamin Doyon}
\affiliation
{Department of Mathematics, King's College London, Strand WC2R 2LS, U.K.
}


\begin{abstract} 
We define a simple model of conformal field theory in random space-time environments, which we refer to as stochastic conformal field theory. This model accounts for the effects of dilute random impurities in strongly interacting critical many-body systems. On one hand, surprisingly, although impurities are separated by macroscopic distances, we find that the infinite-time steady state is factorized on microscopic lengths, a signature of the emergence of localization. The stationary state also displays vanishing energy current and strong uncorrelated spatial fluctuations of local observables. On the other hand, at finite times, the transient shows a crossover from ballistic to diffusive energy propagation.  In this regime and a Markovian limit, concentrating on current-generating initial states with a temperature imbalance, we show that the energy current and density satisfy simple dissipative hydrodynamic equations. We describe the space-time scales at which non-equilibrium currents exist. We show that a light-cone effect subsists in the presence of impurities although a momentum burst propagates transiently on a diffusive scale only.
\end{abstract}

\maketitle

\section{Introduction}

One of the most interesting and active areas of current research is the physics of randomness in extended, interacting quantum systems. Since the work of Anderson \cite{Anderson}, it is well known that the effects of random external fields and impurities may have important physical consequences, such a vanishing conductivity and the absence of ergodicity \cite{Localisation}. The situation is more complex in extended systems where strong interactions play a crucial role, leading to many-body localization (MBL) \cite{Many-Body}.

In this letter, we ask the related question of the interplay between emergent collective behaviours in interacting quantum critical systems and randomness due to dilute impurities. On the one hand, perhaps the most striking effect of strong interactions in extended systems is the emergence near criticality of {\it large-scale collective behaviours} and large-distance correlations, with associated {\it ballistic propagation} of energy. On the other hand, in many-body localized states local degrees of freedom become largely uncorrelated, precluding transport. Can dilute impurities strongly affect collective behaviours and break ballisticity? May signatures of localization emerge? To our knowledge these questions have not been addressed yet in the literature.

Conformal field theory (CFT) is the most powerful theory for emerging collective behaviour at critical points \cite{CFT}, hence the foremost playground for understanding new many-body physics. We introduce the concept of stochastic CFT, and we use it in order to gain insight into the physics of diffusion and the emergence of localization in one-dimensional quantum critical systems.  Stochastic CFT, inspired by open quantum Brownian motion \cite{OQBM}, is the combination of fundamental notions of CFT such as ballistic energy transport, with randomness in space and time. It can be interpreted as an effective model for the interplay of many-body collective behaviours with dilute, randomly placed impurities. It explicitly describes quantum evolution in a random environment of classically fluctuating impurities.

Interestingly, dynamics in this model displays the passage from ballistic, to diffusive, and then to localized physics. For this purpose, we show how some of the important tools of dissipative quantum mechanics \cite{Breuer} can be extended to CFT. This provides a precise notion of diffusivity, which is in line with that used in recent works on quantum chains \cite{diff0} (but which does not necessarily correspond to de Gennes' phenomenological theory \cite{diff}). Surprisingly, we also show that after a long time an important signature of localization emerges: we find that the stationary state of {\it macroscopically} separated random impurities in interacting, critical many-body systems, is correlated only at {\it microscopic} length scales. Randomness (or more precisely irregularity) is important: regularly placed and fluctuating impurities would not give rise to microscopic correlation lengths. This gives an analytically tractable framework in which to extract certain principles at the roots of such localization effects in interacting systems.

\section{Random impurities and stochastic CFT} \label{sectII}

{\it Heuristics.}
Consider a quasi-one-dimensional metal at low energies. The Luttinger liquid theory predicts that the electronic behaviour is described by a one-dimensional CFT. As is well known, this supports ballistic transport of various quantities such as the energy, and thus an infinite conductivity.

In real, non-ideal, systems, electrons are subject to diffusive processes due to scattering on impurities or with other degree freedoms, and this modifies the dynamics. Two types of scattering may occur: elastic, inducing momentum, but no energy, relaxation, or inelastic, inducing energy relaxation and phase decoherence. In the mesoscopic regime the mean free path is much smaller than the decoherence length so that coherence effects are still important, and the system is described by elastic scattering events separated by ballistic propagations. We concentrate on this regime.

At low energies and low defect densities, any physical impurity will be described by a conformal defect. Consider the local observables representing conserved currents, such as the energy current. These are chiral: despite the strong interaction between electrons, the local energy is ballistically transported by independent right and left-moving collective energy packets. The associated local observables are the components $T^+(x) = (h(x) + p(x))/2$ and $T^-(x) = (h(x)-p(x))/2$ of the stress-energy tensor, where $h(x)$ is the energy density and $p(x)$ is the momentum density.

The model we consider is that where defects are positioned randomly in space, and also ``on'' or ``off'' in a random fashion in time. This is the simplest model of a {\it classically fluctuating environment}: the quantum system is connected to an environment only via sparse impurities, activated and deactivated in a random fashion due to classical fluctuations. We expect the model mimics the quasi-classical cumulative effect of random interference that would occur upon scattering off fixed, time-independent impurities with nontrivial reflection and transmission. One may further generalize the model to random velocity fields. CFT with randomly placed, stochastic conformal defects or random velocity fields is what we refer to as {\it stochastic CFT}.
\begin{figure} \label{fig:fig1}
\bc\includegraphics[width=5cm]{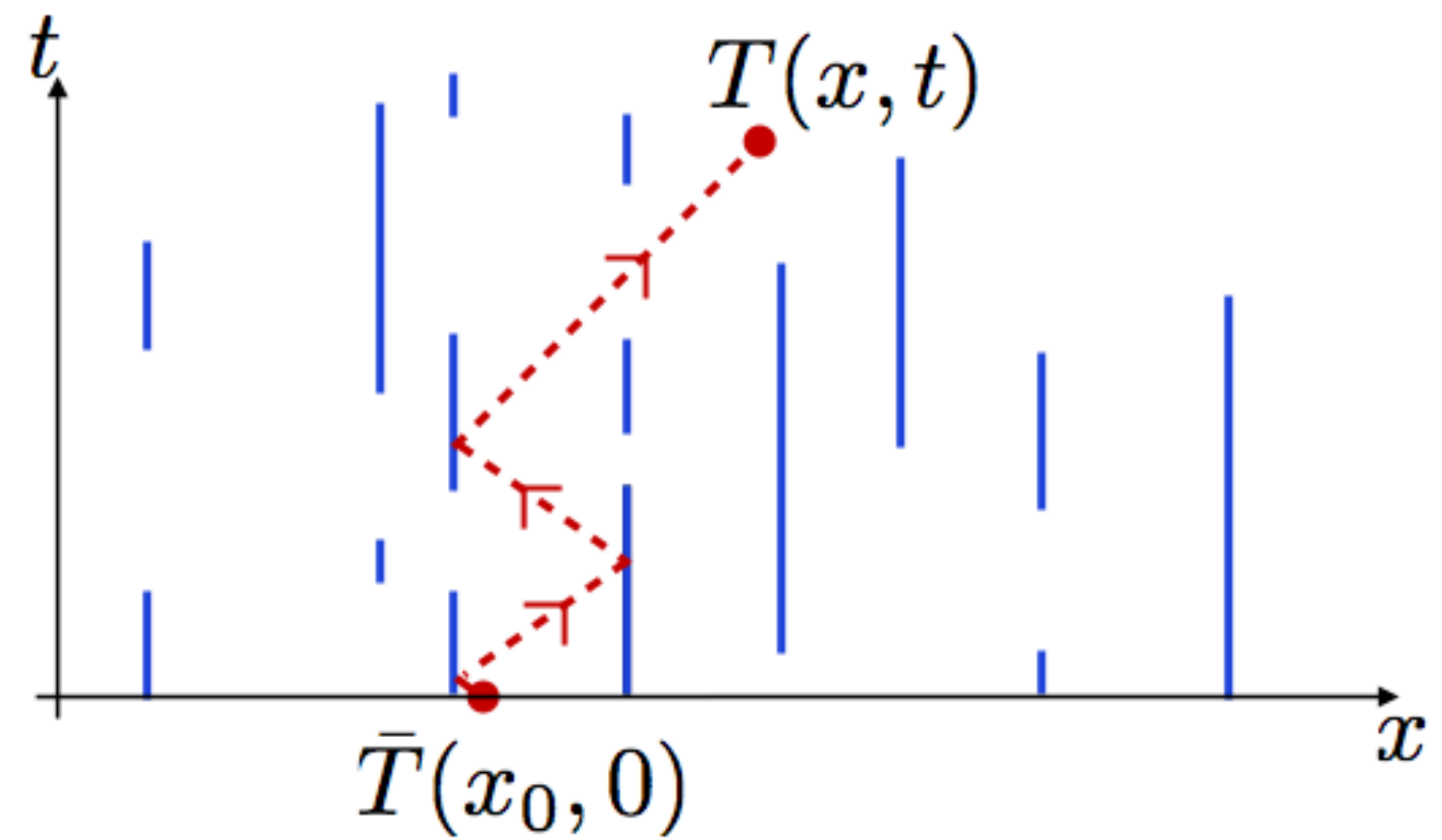}\ec
\caption{A pictorial representation of a sample of the random space-time field of conformal defects (the vertical segments), with the associated propagation trajectory of $T(x,t)$ (the dotted line); eq. \eqref{Utrajs}, in this case with $\ep=+$ and $\ep_0=-$.}
\end{figure}

Whereas introducing stochastic impurities aims at mimicking destructive interferences, 
the use of CFT allows us to extract the relevant collective modes in the spirit of the renormalization group. The emergence of conformal invariance at low energy provides a universal description of transport phenomena in these critical systems encompassed in the chiral nature of energy transport. It ultimately leads to the description via random trajectories given below. It applies to interacting systems as our model deals with CFT with arbitrary central charge. Note however that although appropriate to critical energy transport in the presence of impurities, this description does not provide a direct way to disentangle the effect randomness on non chiral observables, which are typically present in interacting models.

{\it Time evolution and random trajectories.}
Many of the conclusions below are valid under quite general setup for stochastic CFT. However, in order to be precise, it is convenient to specify a model of stochastic CFT.

We place defects uniformly in space $\R$, and each defect is activated and deactivated in a random fashion in time, independently of each other. The random set of positions $X\subset \R$ is a Poisson process of intensity $2\eta$, so that the probability of finding a defect between $x$ and $x+\dd x$ is $\prob([x,x+\dd x]\cap X\neq\emptyset) = 2\eta \dd x$. For each $x\in X$, the random set of activation / deactivation events $\{t^x_j,\,j=0,1,2,\ldots\}\subset \R^+$ (with $t_0=0)$ is an independent Poisson process of intensity $\nu$. The description starts at time $t=0$,  and each defect starts either in an activated or deactivated state with equal probabilities, $b^x\in\{0,1\}$ with probabilities $1/2,1/2$. We denote by $T^x$ the union of all activation time intervals for the defect at position $x$; this is $T^x = \cup_{k=0}^{\infty} [t^x_{2k+b^x},t^x_{2k+1+b^x}]$. We also denote by $\mathfrak{D} :=\{(x,t):x\in X,\,t\in T^x\}$ the subset of space-time $\R^2$ where an activated defect lies. This is a union of infinitely-many finite vertical segments lying in the upper half plane (which definiteness we consider closed). As an example, let $p(x,\delta)\,dx$ be the probability that a segment intersects some horizontal line $t=\mathrm{const.}$ between $x$ and $x+\dd x$, and lasts for a time longer than $\delta$ after $t$ (i.e. it ends at a time bigger than $t+\delta$). One can show that
\beq\label{pcont}
	p(x,\delta)\,\dd x = \eta e^{-\nu \delta}\dd x.
\eeq

Given a sample $S$ of random defects, there are associated time-dependent evolution operators $U_{t;s}$ describing evolution of states from any time $s$ to $t>s$ within this defect configuration. By construction they satisfy $U_{t;s}U_{s;0} = U_{t;0}$ 
and the time-evolved observables are $\Or(x,t) = U_{t;0}^\dag \Or(x) U_{t;0}$. Averages in the time-evolved state $\bra\cdots\ket_t$ are obtained, in the Heisenberg picture, as averages of time-evolved observables in the initial state, $\bra \Or\ket_t := \bra U_{t;0}^\dag\,\Or\, U_{t;0}\ket_0$. 

Since $\mathfrak{D}$ is random, so is $U_{t;s}$ for any $t>s$. Since $U_{t;s}$ represents evolution from time $s$ to time $t$, it is completely determined by the defects that are active within that time period. Let $H_t$ be the Hamiltonian representing instantaneous evolution within the defect configuration exactly at time $t$. Then $\ii\p_tU_{t;0} = H_t U_{t;0}$ and therefore $\p_t \Or(x,t) = \ii\, U_{t;0}^\dag [H_t,\Or(x)] U_{t;0}$.
The vector space spanned by right/left-moving densities $T^\ep(x)$ ($x\in\R$ and $\ep\in\{\pm\}$) is the space of observables on which we wish to describe time evolution.  For every $x$ that lies away from an active defect at time $t$, that is $x\in \R: t\not\in T^x$, we have chiral evolution, $\ii [H_t,T^\ep(x)] = -\ep\p_x T^\ep(x)$ (here and below, the Fermi velocity is set to unity). Thus we find $\p_t T^\ep(x,t) + \ep \p_x T^\ep(x,t) = 0$ for all  $(x,t)\in \R^2\setminus \mathfrak{D}$. That is, the time evolved operator $T^\ep(x,t)$ is obtained by drawing a trajectory starting from $(x,t)$ and going backward in time towards the left (right) if $\ep>0$ ($\ep<0$), as long as this diagonal does not intersect any active impurity.

On both sides of a defect, we impose conformal boundary conditions. On the energy sector, a conformal boundary condition at the point $x_0$ is the operator equality $T^\ep(x_0) = T^{-\ep}(x_0)$. It implies that, on both sides of the defect, the trajectory performs a reflection against the defect: we map $T^\ep(x_0)\mapsto T^{-\ep}(x_0)$ and then continue the chiral backward evolution using the new, reflected operator, which will then go away from the defect. This is repeated as many times as active impurities are hit, until time 0 is reached, where the resulting operator can be evaluated within the initial state. See Fig.1.

The full time evolution is therefore described in terms of trajectories: for every $\mathfrak{D}$ and every $x,t$, the  trajectory
\beq \label{eq:trajec}
	s\in[0,t]\mapsto (\ep_s,x_s),\quad \ep_t=\ep,\,x_t=x
\eeq
is obtained by evolution backward from $(\ep,x)$ at time $t$, chiral between defects and reflecting on defects. Time evolution is described by the transport equation:
\beq\label{Utrajs}
	T^\ep(x,t) = T^{\ep_s}(x_s,s),\quad \forall\; s<t.
\eeq
Taking $s=0$, we have $T^\ep(x,t) = T^{\ep_0}(x_0)$.
Therefore,
\beq\label{UTexpect}
	\Big\bra \prod_{j=1}^N T^{\ep^{(j)}}(x^{(j)})\Big\ket_t
	=
	\Big\bra \prod_{j=1}^N T^{\ep^{(j)}_0}(x^{(j)}_0)\Big\ket_0.
\eeq
Since time evolution is random, the quantum average at time $t$ is also a random value. The measure on $\mathfrak{D}$ induces a probability measure on the trajectories.

We therefore need to evaluate the stochastic expectation of random operators,
\beq \label{CPmap}
	\Phi_{t;s}^*[\Or]:=\expec \Big[U_{t;s}^\dag
	\Or\,U_{t;s}\Big].
\eeq
In general, $\Phi^*_{t;s}$ is {\it not} a unitary transformation. However, if $U_{t;s}$ are unitary, it is a completely positive (CP) map and it defines a {\it dissipative dynamics} \cite{CP-evol,Breuer}. It can be defined on density matrices by duality. 
Thanks to \eqref{Utrajs}, we may express the map $\Phi_{t;s}^*$ on product of stress tensor components in terms of averages over trajectories:
\beq\label{PhiT}
	\Phi_{t;s}^*\Big[\prod_{j=1}^N T^{\ep^{(j)}}(x^{(j)})\Big]
	= \expec\Big[ \prod_{j=1}^N T^{\ep^{(j)}_s}(x^{(j)}_s)\Big]
\eeq
where on the right-hand side, the expectation is over the trajectories ending at $(\ep^{(j)},x^{(j)})$ at time $t$.

Inverting the order of the quantum and stochastic averages, we evaluate the combined average in the time-evolved state as follows:
\beq
	\expec \lt[\bra \Or\ket_t\rt]
	=
	\bra\, \Phi_{t;0}^*[\Or] \,\ket_0,
\eeq
which can be evaluated using eqs. (\ref{UTexpect},\ref{PhiT}).

We make three remarks, which are further developed in the Supplementary Material (SM):
(i) The evolution (\ref{Utrajs}) is almost surely unitary on products of stress tensor components. The events, of measure zero, on which unitary is broken correspond to cases in which two infinitely close nearby trajectories split because one hits an activated impurity while the other does not. The physical explanation of this effect is that at the microscopic level, the operations of activation and deactivation are high-energy operations, involving states that do not lie in the low-energy region of the spectrum described by CFT. Non-universal, high-energy effects are therefore involved that are not encoded by the simple activation / deactivation of conformal defects. However, since these events are extremely rare, they do not affect the results, which are thus universal. (ii) The dynamics  (\ref{CPmap}) is non-Markovian because the time evolution operators $U_{t;s}$ and $U_{s;u}$ (for $t<s<u$) are not independent, but it becomes Markovian in the limit $\nu^{-1}\to 0$.  (iii) The dynamics (\ref{Utrajs}) is a transport equation and the above construction can be generalized to other random fields.

\section{Signatures of localisation at infinite time}

The steady state is naturally defined as the large-time limit of the stochastic average of the quantum state,
\beq
	\bra \Or\ket_{\rm stat} := \lim_{t\to\infty} \expec\big[\bra \Or\ket_t\big].
\eeq
Recall that the stochastic average mimics the many interference effects of the large number of partial transmission and reflection events on impurities. It is possible to provide rather general arguments for the physical properties of the steady state arising, at large times, from stochastic CFT.  The main results are as follows.

We consider an initial state, possibly inhomogeneous, with well defined asymptotics in the far right and left,
\[
	\Big\bra \prod_{j=1}^N T^{\ep^{(j)}}(x^{(j)})\Big\ket_{\rm r,l} := \lim_{x\to\pm\infty}\Big\bra \prod_{j=1}^N T^{\ep^{(j)}}(x^{(j)}+x)\Big\ket.
\]
Assume that each initial asymptotic state $\bra\cdots\ket_{\rm r,l}$ is clustering at large distances. These are translation invariant, we denote the one-point functions as $\bra T^\ep\ket_{\rm r,l}:=\bra T^\ep(x)\ket_{\rm r,l}$.

{\it Microscopic clustering.} The steady state is factorized on microscopic scales because backwards random trajectories almost surely split far apart. Since small time evolution is almost surely ballistic, factorization also occurs in time-dependent correlations, a fortiori at larger times as well. Formally, $\bra\Or\ket_{\rm stat} = \prod_{x\in\R} \bra \Or|_x \ket_{\rm micro}$
where $\Or|_x$ is the component of $\Or$ supported at the position $x$:
\beq
	\Big\bra \prod_{j=1}^N T^{\ep^{(j)}}(x^{(j)},t^{(j)})\Big\ket_{\rm stat}
	=\prod_{j=1}^N \bra T^{\ep^{(j)}}(x^{(j)})\ket_{\rm micro}.
\eeq
Because backwards trajectories have equal probability to be on the left or right and to be either left or right-moving types, the microscopic-scale state $\bra \cdots \ket_{\rm micro}$ is given by
\beq
	\bra T^\ep(x)\ket_{\rm micro} =\frc14 
	\sum_{\ep'\in\{\pm\}\atop \eta\in\{{\rm r,l}\}}
	\bra T^{\ep'}\ket_\eta.
\eeq

Microscopic clustering indicates that the state obtained at large times is the {\it scaling limit of a localized state}: the steady state has a correlation length that is zero in the renormalized distances of CFT, thus finite in units of lattice spacing; and clustering also holds under time separation. Microscopic clustering occurs even though impurities are separated by {\it macroscopic} distances: randomness at the macroscopic scale makes correlations microscopic. This occurs independently from the correlation properties of the initial state (under the above weak conditions). These are strong signatures of localization of the stationary state. Note that in quantum states with microscopic correlation length but with ballistic components in transport, the real-time current correlations are large on light-cone rays; this ballistic effect is not seen in our model further pointing to absence of transport in the steady state. Note also that if impurities were distributed in space-time in a regular fashion, trajectories would not almost surely split apart, and thus microscopic clustering would not occur.

{\it Vanishing transport.} It is clear from the above results that the mean energy current is zero at every point, $\bra p(x)\ket_{\rm stat} = 0$.
This is true no matter the properties of the initial state: it may be a non-equilibrium, current-carrying state with $\bra T^\ep\ket_{\rm r,l}$ dependent on $\ep$, or a spatially unbalanced state with $\bra T^\ep\ket_{\rm r,l}$ dependent on ${\rm r,l}$. Another related proposition holds. One may look not only at the stochastic average of the quantum states, but also at the stochastic fluctuations. These represent, within the stochastic CFT modelling, quantum fluctuations occurring at large times due to interference effects. It is possible to show that the current is zero {\it on spatial average}, almost surely at large time, $\int_{a}^b \dd x\,\bra p(x)\ket_{t\to\infty} = 0$. Yet, the local currents are far from being zero before the stochastic averages, and in particular the integration of the squares of the quantum-average momenta is nonzero: $\int_{a}^b \dd x\,\big(\bra p(x)\ket_{t\to\infty}\big)^2 >0$ almost surely. The spatially integrated current is self-averaging, but locally very rough:  there are strong local currents within each microscopic cell, and these can be in any direction, independently from one cell to the other.

\begin{figure} \label{fig:fig2}
\vskip 0.2 truecm
\bc\includegraphics[width=6.5cm]{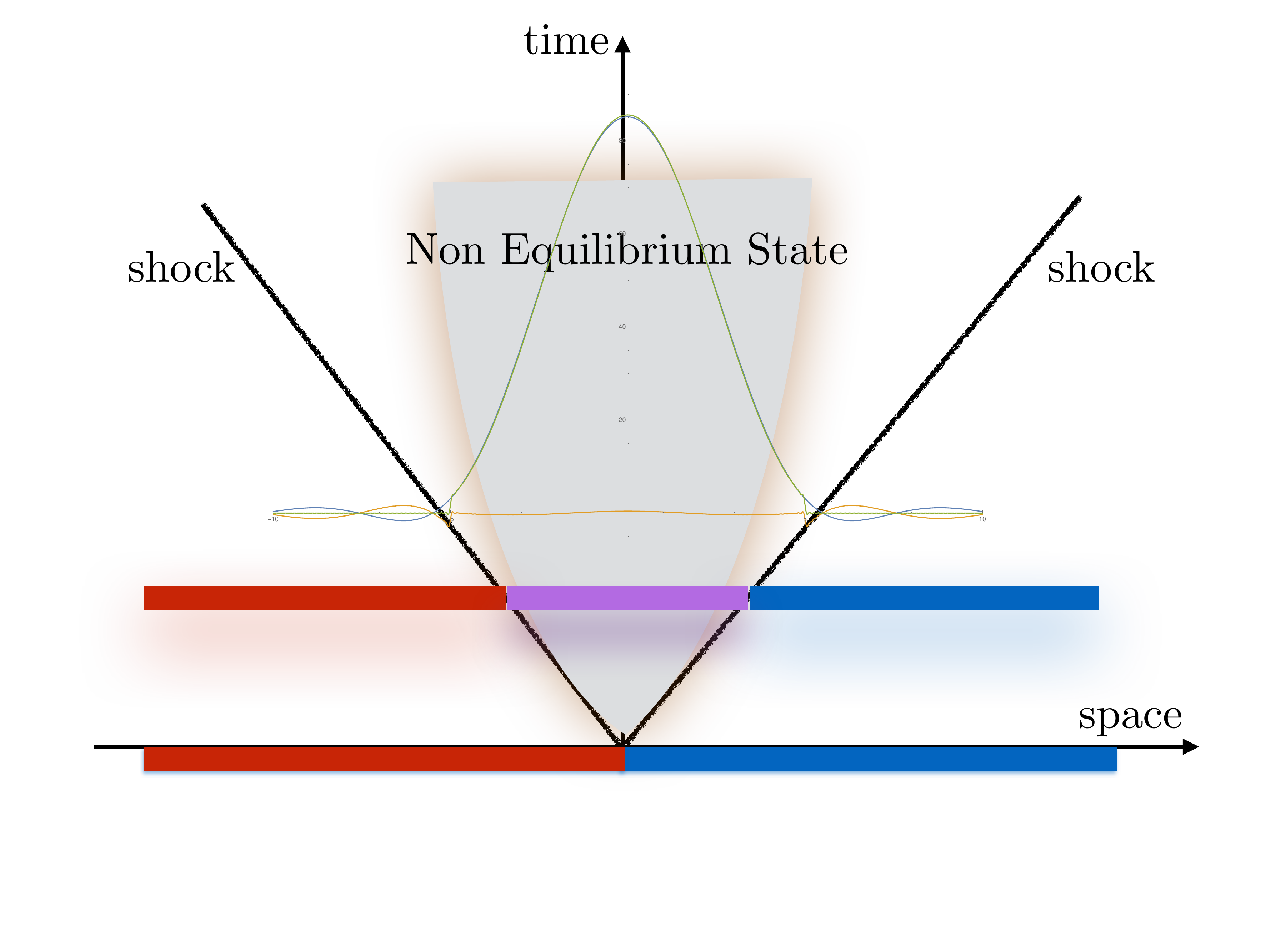}\ec
\caption{The non equilibrium state produced by a localized momentum burst. It is current-carrying inside the light-cone only. The grey region indicates the space-time domain with significant non zero momentum flow. The green  curve is the momentum density at a given time (the blue and yellow curves are its diffusive and ballistic components, respectively).}
\end{figure}

\section{Crossover from ballistic to diffusive transport}

Before reaching the localized stationary state at asymptotically large time, the system does not display localization: rather a transient occurs that is dominated by diffusion. For simplicity consider the asymptotic regime $\nu\gg\eta$, or equivalently when the observation time and the mean impurity spacing become very large simultaneously, $t,\eta^{-1}\to\infty$, $t\eta$ fixed. The dynamics (\ref{CPmap}) becomes Markovian because the impurities are renewed at each time. One may thus expect
\beq \label{eq:dyn-lind}
\partial_t \mathcal{O}(t)= \ii[H_\mathrm{cft},\mathcal{O}](t) + \mathcal{L}_\mathrm{imp}^*(\mathcal{O})(t),
\eeq
with $\mathcal{L}_\mathrm{imp}^*$  the Lindblad operator \cite{Lindblad} generating the dissipative dynamics due to scattering off impurities. 

In this limit, elastic scattering events occur independently at a rate $\eta$. 
It can be shown (see the SM) that the energy and momentum density satisfy the following hydrodynamic-type equations:
\begin{eqnarray} 
\partial_t h(x,t) +\partial_xp(x,t) &=&0, \label{eq:dissipA}\\
\partial_t p(x,t)+\partial_x h(x,t) &=& -2\eta\, p(x,t). \label{eq:dissipB}
\end{eqnarray}
The first equation codes for local energy conservation, characteristic of elastic scattering. The second includes a friction term, due to the scattering events $p(x)\to -p(x)$ at rate $\eta$ and net momentum density deficit of $-2p(x)$. If the momentum density vanishes at infinity (the only case we consider), then the total energy $E:=\int \dd x\, h(x,t)$ is constant in time. The total momentum $P_t:=\int \dd x\, p(x,t)$ exponentially approaches a constant finite value, determined by the energy density at infinity. Eqs. (\ref{eq:dissipA},\ref{eq:dissipB}) were first obtained in the context of the kinetic theory of gases a long time ago \cite{Cattaneo}.

Eqs. (\ref{eq:dissipA},\ref{eq:dissipB}) code for a ballistic-to-diffusive crossover: the evolution is ballistic at short time but diffusive at larger time, see Fig.2. Indeed, they imply
\beq \label{eq:p-seul}
 \partial_t^2 p(x,t) +2\eta\, \partial_t p(x,t) -\partial_x^2 p(x,t) = 0.
 \eeq
At small time the higher-derivative terms dominate, giving the relativistic wave equation $\partial_t^2 p(x,t) -\partial_x^2 p(x,t) = 0$.
At large time, the momentum profile smoothes out and $\partial_t^2 p\ll \eta \partial_t p$, giving the diffusion equation, $\partial_t p(x,t)  = \frac{D_F}{2}\, \partial_x^2 p(x,t)$, with a diffusion constant  $D_F=1/\eta$ \cite{DF}.

The effects are clearly seen in the case where the initial configuration produces an energy carrying non-equilibrium flow. We look at an initial state with unbalanced left (right) energy densities $h_{l}$ ($h_{r})$, as in the non-equilibrium CFT setup \cite{BD-all}. This produces an initial momentum burst, and, in absence of dissipative processes, the state evolves toward a current-carrying steady state with mean energy flow $\frac{1}{2}(h_{l}-h_{r})$. Momentum relaxation modifies this picture.
Eq.(\ref{eq:p-seul}) can be integrated with initial conditions, $p_{t=0}(x)=0$ and $\partial_t p_{t=0}(x)= (h_l-h_r)\, \delta(x)$. It gives $p(x,t)= (h_l-h_r)\, G(x,t)$ with
\beq \label{eq:def-G} 
G(x,t)= e^{-\eta t}\, \int\frac{\dd k}{2\pi}\, \frac{\sin(t\sqrt{k^2-\eta^2 })}{\sqrt{k^2-\eta^2}}\, \cos(kx),
\eeq
with ballistic $(k>\eta)$ and diffusive $(k<\eta)$ components.

This non-equilibrium state is current-carrying inside the light-cone only: $p(x,t)$ is nontrivial inside the light-cone ($p(x,t)\not=0$ for $x^2-t^2<0$), but vanishes outside ($p(x,t)=0$ for $x^2-t^2>0$). This echoes the fact that momentum relaxation does not introduce processes propagating at speeds faster than the Fermi velocity.  Inside the light-cone along the ray $x=vt$, $(v<1)$, the flow decreases exponentially: $p(x=vt,t) \sim \frac{\mathrm{const.}}{\sqrt{\eta t \sqrt{1-v^2}}}\, e^{(\sqrt{1-v^2}-1)\eta t}$. Deep inside the light-cone the diffusive component of the flow is dominating. Near the light-cone the diffusive and ballistic components are of the same order, although both are decaying, and they compensate outside the light-cone. There are contact singularities on the light cone, which are smoothed by irrelevant operators \cite{BD-hydro}. As a consequence, the momentum density has a Gaussian-like bell profile with width of order $\sqrt{t}$, much smaller than the ballistic propagation length $t$: a momentum burst $\partial_t p_{t=0}(x)\propto \delta(x)$ propagates  diffusively only. On the ballistic scale of order $t$, the momentum flow remains concentrated at the origin.

\section{Conclusion}

We have defined a model of stochastic conformal field theory as an effective way to deal with quantum interferences due to multiple elastic scattering on impurities in one-dimensional critical extended interacting many-body system. We have shown that (1) the steady state is localized, and (2) the transient is formed by a crossover from ballistic to diffusive transport due to elastic momentum relaxation. We thus find, over time, the passage from ballistic to diffusive and then to localized physics. Note that it was recently argued, based on a single-particle picture, that ``continuous systems" delocalize \cite{GMMP}; however this does not apply to the class of emergent collective behaviors described by CFT. 
\bigskip

{\it Acknowledgments}: We thank John Cardy, Sebasti\'an Montes, Piero Naldesi, Javier Rodr\'iguez and German Sierra for discussions. This work is supported in part by the ANR project ``StoQ'' n${}^\circ$ANR-14-CE25-0003.


\vfill \eject


\section*{Supplementary Material}

\underline{Stochastic CFTs as effective theories.}

This first set of remarks aims at making clear within which context our work has to be perceived. They arose while answering the following questions asked by the referees: Why our approach deals with interacting systems? What is the role of interaction or lack thereof? How and why a seemingly complicated problem reduces to the study of right turning/left turning trajectories (where normally one would consider interacting trajectories, two-particle states, three particle states etc.)? What is the significance of the fact that the analysis boils down to the study of the chaotic trajectories?

The understanding of many physical phenomena, ranging from fundamental interactions to classical or quantum extended systems, requires extracting the relevant large-scale degrees of freedom, which can manifest themselves in different guises, say collectives modes, shapes, structures or variables. Extracting these relevant variables is of course one of the main aim of the renormalization group, but making the RG work often requires having recognized the appropriate setup (in the opposite case, one cannot make the RG transformations contracting towards a relevant effective model, and as an illustration, the absence of such appropriate setup is for instance why the RG has not been yet successfully applied to turbulence). For 1D gapless many-body systems the situation is much more favorable and much better controlled because, in one hand, the RG transformations do converge towards low energy fixed points that are describable by conformal field theories and, on the other hand, conformal symmetry in 1+1D is powerful enough to disentangle the dynamics of observables in the energy sector. All these facts are well known and we use them to claim that the effective model we introduced deals with the effective theories of interacting systems at low energy.  We believe that this explains how and why a seemingly complicated problem reduces to the study of right turning/left turning trajectories.

These trajectories are those of collective modes --the relevant low energy modes-- which, for strongly interacting systems, share only very few characters with the original microscopic modes of the system models. In other words, by concentrating on those modes and their trajectories we have extracted the relevant variables which disentangle the dynamics (in a way similar as one does with action-angle variables in complicated integrable models; here the original microscopic model need not be integrable, even though the effective low energy theory is).  The universality of this dynamics at low energy is encoded into the algebra of observables in the energy sector. This algebra depends on the central charge which counts the number of effective low energy degrees of freedom (in interacting models, this number is not necessarily an integer) and our results apply to any value of central charge. The new point we have added in our approach is to represent the effects of quantum interference due to scattering on permanent partially transmitting impurities by reflection on {\it intermittent impurities}. One may argue for this effective description as follows : `In CFT, energy packets are not divisible into smaller packets and, as a consequence, the defects preserving the universal low energy conformal description are the purely transmissive and purely reflective; therefore, in this description, energy packets may either reflect off or transmit through a defect, and stochasticity is used to allow for effective partial transmission and reflection'. In other words, we expect that effectively after enough scattering events of energy packets, one may assume that the large amount of interference allows for a description in terms of energy packets to become accurate again. Within such framework, the problem of sparse random impurities becomes a simplified problem of transport in a random fields of obstacles, so that the description in terms of random trajectories is accurate again (that is: these random trajectories of effective modes are expected to be the relevant variables, say in the RG sense, of the present impurity problem). We believe that this makes the reduction down to the study of the chaotic trajectories more intuitive.

Of course, using this description and these relevant variables has some limitations: as pointed out in the main text, it is adapted to the description of observables whose dynamics have an underlying chiral character. In fact this is an aspect that intrinsically distinguishes interacting from non-interacting models: in the latter, fields representing the fundamental particles are themselves chiral, hence should be subject to similar random trajectories; in the former, this is not the case, and our theory does not give predictions for such observables. It is by concentrating on the energy sector that such strong results, with such a simple description, is possible even in interacting models. We think this explains what is the role of interaction or lack thereof, and how large is the class of systems for which one should expect this to occur. In particular, we repeat that the results apply to every 1D gapless many-body systems with sparse impurities.


\bigskip

\underline{\it Almost sure unitarity and non-universal effects.}

The activation and deactivation of a conformal impurity in CFT are operations that lead to {\it non-unitary time evolution} on rare events due to short distance (UV) effects. Technically, this is seen as follows. For a given defect configuration $\mathfrak{D}$, let a trajectory $s\mapsto (\ep_s,x_s)$, eq. (2) in the main text, which we will refer to as an $(\ep,x,t)$-trajectory, cross an activation time of an impurity at space-time point $(y,u)$ (an not other activation or deactivation time at later times $s\in[u,t]$). Since activation / deactivation times are, according to our choice, times when the impurity is activated, then this trajectory will reflect. Consider an $(\ep,x+\delta x,t)$-trajectory $s\mapsto (\ep_s',x_s')$, for $|\delta x|$ small enough. Away from impurities, it stays parallel to the $(\ep,x,t)$-trajectory for times $s\in[u,t]$. However, it either will hit the activated impurity at time $s=u$ if $\delta x>0$ and miss it if $\delta x<0$, or the opposite. If the impurity is missed, then the $(\ep,x,t)$-trajectory and $(\ep,x+\delta x,t)$-trajectory will not stay parallel away from impurities at times $s\in[0,u]$; they are uncorrelated at these times. As a consequence, at time 0, the positions $x_0$ and $x_0'$ will be (almost surely) a finite distance from each other, either for $\delta x>0$ or for $\delta x<0$. Therefore, the leading term in the operator product expansion (OPE) $T^\ep(x,t)T^\ep(x+\delta x,t) = T^{\ep_0}(x_0)T^{\ep'_0}(x'_0)$ is not singular as $\delta x
\to0$, either for $\delta x>0$ or $\delta x<0$. Hence it is {\it not} that induced from a unitary transformation of the standard OPE $T^\ep(x) T^\ep(x+\delta x) \propto (\delta x)^{-4} {\bf 1}$. 

The physical explanation of this effect is that at the microscopic level, the operations of activation and deactivation are high-energy operations, involving states that do not lie in the low-energy region of the spectrum described by CFT. Non-universal, high-energy effects are therefore involved that are not encoded by the simple activation / deactivation of conformal defects. As a consequence, the CFT evolution is a effective on the low-energy degrees of freedom, and unitarity is partially lost because of the lost of information on the high-energy degrees of freedom. Nevertheless, the action $U_{t;s}^\dag \cdots U_{t;s}$ on any finite product $\prod_{j=1}^N T^{\ep^{(j)}}(x^{(j)})$ is {\it almost surely} a unitary action with respect to the measure on $S$, as almost surely the trajectories will miss any activation / deactivation space-time points within the time interval between $s$ and $t$. Since stochastic CFT is a model for random impurities where part of quantum randomness has been replaced by stochasticity, it is meaningful after averaging, under which only almost sure properties subsist, and thus these non-unitary effects do not affect the physical conclusions from this model.
\bigskip

\underline{\it  Non-Markovianity versus Markovianity.}

 If the unitary evolution operators $U_{t;s}$ and $U_{s;u}$ (for $t<s<u$) were independent random variables -- that is, if the random Hamiltonians $H_t$ and $H_{t'}$ were independent for any $t\neq t'$ -- then two important conclusions would follow: (1) the random $(\ep,x,t)$-trajectories would be Markovian, and (2) the CP maps $\Phi_{t;s}$ would factorize as $\Phi_{t;0}= \Phi_{t;s}\circ \Phi_{s;0}$. Indeed, the evolution, backward in time, of a trajectory from a point $(\ep_u,x_u)$ at time $u$, obtained using $T^{\ep_u}(x_u,u) = U_{u;0}^\dag T^{\ep_u}(x_u)U_{0;u}$, is entirely determined by the value of $\ep_u$ and $x_u$ and by the probability measure on $U_{u;0}$. Thus in the case where such measures factorize, it would be independent of the shape of the part of the trajectory already traced, $(\ep_s,x_s) :s\in[u,t]$. Similarly, the expectation in (5) (main text) would factorize, leading to the composition property of the CP maps. This latter property is important for many calculations, as it means that these maps define a one-parameter semi-group.

However, it turns out that the unitary evolution operators $U_{t;s}$ and $U_{s;u}$ (for $t<s<u$) are {\it not} independent random variables. For instance, the probability that a segment covers the time interval $[t,s]$ at position $x$ knowing that there was an active impurity at space-time point $(x,s)$ is $e^{-\nu (t-s)}$, and is much greater than the probability of the same event under the knowledge that there was no active impurity $(x,s)$, which is proportional to $\eta\dd x$. As a consequence, the random trajectories are not Markovian, and the CP maps $\Phi_{t;s}$ do not factorize, $\Phi_{t;0}\neq \Phi_{t;s}\circ \Phi_{s;0}$.

Yet, there is a limit where Markovianity is recovered: the limit where the mean activation / deactivation time is very small as compared with the mean distance between impurities. This is obtained by taking $\nu\gg\eta$, or equivalently by taking the observation time and mean impurity spacing very large simultaneously, $t,\eta^{-1}\to\infty$, $t\eta$ fixed. Indeed, in this limit, it is overwhelmingly less likely that a trajectory hits an impurity twice on the same activation period, than it is that it hits it once. Therefore, collision events become independent, and we may consider each impurity as being ``semi-transparent'', with trajectories transmitting through or reflecting with equal probability independently at each collision event. In this limit, trajectories are Markovian, and the CP maps $\Phi_{t;s}$ form a one-parameter semi-group. This limit is a good tool in order to analyse large-time behaviours, which we will make use below.

Let us finally give a simple derivation that the dynamical CP maps form a one parameter group if Markovianity holds, that if evolution operator $U_{t;s}$, $(t>s)$, and $U_{s;0}$ are independent. Recall the definition (\ref{CPmap}) of the dynamical CP-maps, $	\Phi_{t;s}^*[\Or]:=\expec [U_{t;s}^\dag\Or\,U_{t;s}]$. By time translation invariance of the measure, $\Phi_{t;s}^*$ only depends on the time difference $t-s$. Then, 
\beqs
\Phi^*_{t;0}[\Or] &=&  \mathbb{E}[ U_{t;0}^* \Or U_{t;0}] \\
&=&  \mathbb{E}[ U_{s;0}^* U_{t;s}^*\Or U_{t;s}U_{s;0}],\quad \mathrm{by}\ U_{t;0}=U_{t;s}U_{s;0}, \\
&=&  \mathbb{E}[ U_{s;0}^* \mathbb{E}[U_{t;s}^*\Or U_{t;s}]U_{s;0}], \quad \mathrm{by\ independence},\\
&=& \Phi^*_{s;0}\circ \Phi^*_{t;s}[\Or].
\eeqs
If Markovianity does not hold, we may rescue some of the property of the composition law by considering an alternative definition of the CP-maps as the conditioned expectation $\expec [U_{t;s}^\dag\Or\,U_{t;s}\vert \mathcal{F}_s]$ with $\mathcal{F}_s$ the filtration up to time $s$.

\bigskip

\underline{\it Random velocity fields and transport.}

It is possible to generalise the above construction to other random fields. For instance, in a random metric, the time-evolution equation is that in a random velocity field $u^\ep(x,t)$:
\beq
 \partial_t T^\ep(x,t) + u^\ep(x,t)\partial_x T^\ep(x,t)=0. 
 \eeq
This leads to natural $(\ep,x,t)$-trajectories along the field,
\[ \p_s x_s = u^\ep(x_s,s),\quad x_t = x.\]
One may likewise consider random ``acceleration field'', which will likewise affect the parameter $\ep$
\beqs 
\partial_t T^\ep(x,t) &+& u^\ep(x,t)\partial_x T^\ep(x,t) \\
	&+&  a^\ep(x,t) (T^{-\ep}(x,t)-T^{\ep}(x,t))=0. 
\eeqs
 The random impurity fields we have described is a pure delta-function acceleration field, implementing random hard reflections. Acceleration fields will lead, generically, to non-unitary effects.
  
 \bigskip
 
 \underline{\it Markovian limit and a Lindblad-like operator.}
 
In the limit $\nu\to\infty$, it is clear that the stochastic process (\ref{eq:trajec}) for random trajectories is Markovian (because in effect, the positions of active impurities position are renewed at every time), and hence the CP maps (\ref{CPmap}) form a one parameter semi-group. At the same time, since in this limit each collision event is effectively at an activation / deactivation point, unitarity may be partially lost. Temporarily disregarding potential unitarity problems (but see below), Markovianity of the stochastic process implies that trajectories are generated by a Lindblad operator and the dynamics reads as in eq. (\ref{eq:dyn-lind}).

A simple argument for the form of the Lindblad operator on the energy and momentum densities is as follows. Because nontrivial elastic scattering in one dimension consists in reflecting the momentum, the Lindbladian preserves the energy density $h(x)$ but reverses the sign of the momentum density  $p(x)$. Thus we shall have 
\beq
 \mathcal{L}_\mathrm{imp}^*(h(x))=0,\quad \mathcal{L}_\mathrm{imp}^*(p(x))= - 2\eta\, p(x),
  \eeq
with $\eta$ the collision rate. This clearly codes for the momentum relaxation with transition from $p(x)\to -p(x)$, with a net momentum density deficit of $-2p(x)$, at a rate $\eta$. It yields eqs. (\ref{eq:dissipA},\ref{eq:dissipB}).

A direct derivation, along with an explicit expression for the Lindblad-like operator, can also be obtained. Let $\sigma$ be the operator exchanging left and right movers. In particular, it exchanges the two chiral components of the CFT stress tensor: $\sigma T \sigma = \b T$ and $\sigma \b T \sigma = T$. This operator can be identified with the anti-unitary time-reversal operator, and satisfies $\sigma^2=\mathbb{I}$. Consider the time evolution of a local field $\Or$. At every instant this field is either unitarily evolved with $e^{\ii \dd t \ad H}$ (where $H$ is the hamiltonian without impurity), or reflected with $\Ad \sigma$. Dividing time into small slices of periods $\dd t$, the stochastic process can be described by a series of flips that may occur at times $t=m\dd t$, $m=0,1,2,\ldots$, separated by small unitary evolutions of time $\dd t$:
\beq\label{tevol}
	\Or(t) = \prod_{m=0}^{M} \lt(
	e^{\ii \dd t \ad H} (\Ad \sigma)^{{\tt N}(m\dd t)}\rt)
	\big(\Or\big)
\eeq
where the product is ordered by increasing values of $m$ from the right to the left. The variables ${\tt N}(m\dd t)\in\{0,1\}$ for $m=0,1,2,\ldots$ are i.i.d random variables, indicating if a flip occurred or not. Since the density of impurities in space is $2\eta \dd x$, and since each impurity is activated with probability $1/2$, the random variables ${\tt N}(m\dd t)$ have the distribution
\beqs
	\prob({\tt N}(t)=1) = \eta \dd t.
\eeqs
Therefore, the average of $\Or$ satisfies:
\beqs
	\lefteqn{\frc{\mathbb{E}(\Or(t+\dd t)) - \mathbb{E}(\Or(t))}{\dd t}} &&\n
	&=& \mathbb{E}\lt(\frc{e^{\ii \dd t \ad H} (\Ad \sigma)^{{\tt N}(t)}-1}{\dd t}\,(\Or(t))\rt) \n
	&=& \mathbb{E}\lt(\frc{e^{\ii \dd t \ad H} (\Ad \sigma)^{{\tt N}(t)}-1}{\dd t}\rt)\,\big(\mathbb{E}(\Or(t))\big) \n
	&=& \Big[\eta\dd t\lt(\frc{e^{\ii \dd t \ad H} \Ad \sigma-1}{\dd t}\rt)
	+ \n &&\qquad +\ (1-\eta\dd t)\lt(\frc{e^{\ii \dd t \ad H}-1}{\dd t}\rt)
	\Big]\big(\mathbb{E}(\Or(t))\big) \n
	&=& \big[\eta(\Ad \sigma-1)
	+ \ii \ad H + O(\dd t)
	\big]\,\big(\mathbb{E}(\Or(t))\big)
\eeqs
where in the second step we used Markovianity. This gives (\ref{eq:dyn-lind}) with
\beq
 \mathcal{L}_\mathrm{imp}^*(\mathcal{O})= \eta\, ( \sigma\, \mathcal{O}\, \sigma - \mathcal{O}).
 \eeq
The left-right flip at rate $\eta$ is here clearly apparent.

Although intuitive and precise, there are two difficulties with this approach: (1) the Lindblad equation (11) (main text) with the above Lindbladian is only valid for $\Or$ being a {\it local} field (the stress tensor components or their descendants), not a product of local fields at different positions; and (2) the operator $\sigma$ is {\it not} linear, it is rather anti-linear. There exist no well-defined self-adjoint linear operator in CFT whose action on the chiral stress tensor components is that of $\sigma$ (exchanging $T$ and $\b T$); and on product of many local fields, a different operator than $\sigma$ would have to be used in order to correctly reproduce the Markovian process. These obstructions in constructing a Lindbladian in the usual form are linked to the breakdown of unitary discussed above.

Using the random-flip picture, it is possible to obtain a general solution to the Lindblad-like equation. We can re-arrange the order of the operators in \eqref{tevol} in order to have all $\Ad \sigma$ operators on the right and all $e^{\ii \ad H}$ operators on the left. This is done using $\Ad \sigma \,e^{\ii \ad H} =  e^{-\ii \ad H} \Ad\sigma$, which follows from $[\Ad \sigma, H]=0$ and the fact that $\sigma$ is anti-unitary. Therefore, we divide the total $\ad H$-evolution time into the total of all evolution periods that have occurred after even flips (with evolution $e^{\ii \tau\,{\rm ad}H}$) and that of all evolution periods that have occurred after odd flips (with time-reversed evolution $e^{-\ii (t-\tau)\,{\rm ad}H}$). We then integrating over the possible even/odd divisions using Poisson distribution,
\beq
	\sum_{n} \eta^n e^{-\eta t}\int_0^t \dd\tau
	\frc{\tau^{n_e}}{n_e!}\frc{(t-\tau)^{n_o}}{n_o!}
	e^{\ii\tau\, \ad H} e^{-\ii(t-\tau)\,\ad H}
	(\Ad g)^n
\eeq
where $(n_e,n_0) = (0,-1),\,(0,0),\,(1,0),\,(1,1),\,(2,1),\ldots$ for $n=0,1,2,3,4,\ldots$ (the first case $n=0$ just kills the integral on $\tau=t$; note that $n_e = [n/2]$). Therefore, we find
\beqa
	\lefteqn{\expec\big[\Or(t)\big] = e^{-\eta t}e^{-\ii t{\rm ad}H}\times}&&\\
	&\times&
	\sum_{n=0}^\infty F_{11}\lt(\lt[\frc n2\rt]+1;n+1;2\ii t\,{\rm ad}H\rt)
	\frc{(\eta t\, {\rm Ad} g)^n}{n!} (\Or)\no
\eeqa
where $F_{11}$ is confluent hypergeometric function.
Applied to $T(x)$, it gives for instance
\beqa
	\lefteqn{e^{\eta t}\expec\big[T(x,t)\big] =T(x-t)} && \n &+&
	\sum_{m=0}^\infty \frc{\eta^{2m+1}}{(m!)^2} \int_0^t \dd\tau\,
	\tau^{m}(t-\tau)^{m}\times \n && \quad\times\,
	\lt(\frc{\eta \tau}{m+1}T(x+t-2\tau) + 
	\b T(x-t+2\tau)\rt)\no
\eeqa
which expresses the average at time $t$ explicitly in terms of the initial condition.

Finally, we note that the Markovian limit has a description in terms of ray optics: any spreading of $(x,t)$ with unit weight over a small space-time region produces, almost surely, a set of trajectories that is completely described by sending a light ray through semi-transparent mirrors at the positions in $X$. In a sense, in the Markovian limit any nontrivial interference pattern is washed out.

\bigskip

\underline{\it A probabilistic proof of the hydrodynamic equations.}

We may also derive the evolution equations (\ref{eq:dissipA},\ref{eq:dissipB}) directly. We simply compare $\expec [T^\epsilon(x,t+\dd t)]$ and $\expec [T^\epsilon(x,t)]$. Recall that $T^\epsilon(x,t)=T^{\epsilon_s}(x_s,s)$ where $s\to(\epsilon_s,x_s)$ is the trajectory ending at $(\epsilon,x)$ at time $t$. Thus 
\beq
 T^\epsilon(x,t+\dd t)=T^{\epsilon_{t;\dd t}}(x_{t;\dd t},t)
 \eeq
 with $(\epsilon_{t;\dd t},x_{t;\dd t})$ the (backward) position of the trajectory at time $t$ ending at $(\epsilon,x)$ at time $t+\dd t$. 
 
 Let us now compute the conditioned expectation $ \expec [T^\epsilon(x,t+\dd t)\vert \mathcal{F}_t] $ with $\mathcal{F}_t$ the filtration up to time $t$. The value of $(\epsilon_{t;\dd t},x_{t;\dd t})$ depends whether there is an active segment in a $dt$-neighbourhood of $x$ at time $t$ lasting at least a time laps $\dd t$. If there is none, then $(\epsilon_{t;\dd t},x_{t;\dd t}) = (\epsilon, x-\epsilon \dd t)$ because there is no reflection; if there is one, then $(\epsilon_{t;\dd t},x_{t;\dd t})= (-\epsilon, x+O(\dd t))$ because the trajectory has been reflected. As a consequence,
\beqs
\mathbb{E}[ T^\epsilon(x,t+\dd t)|\mathcal{F}_t] &=&  \big(1-\mathbb{P}[x,\dd t|\mathcal{F}_t] \dd t\big)\, T^{\epsilon_t}(x_t-\epsilon \dd t,t) 
\\ && + \mathbb{P}[x,\dd t|\mathcal{F}_t]\dd t\, T^{-\epsilon}(x,t) + O(\dd t^2),
\eeqs
where $\mathbb{P}[x,\dd t|\mathcal{F}_t]\dd t$ is the probability, conditioned on $\mathcal{F}_t$, that there be an active segment in a $\dd t$-neighbourhood of $x$ lasting longer than $\dd t$. If Markovianity holds (and only if it holds), then
\[ \mathbb{P}[x,\dd t|\mathcal{F}_t]\dd t = p(x,\dd t)\dd t=\eta \dd t+O(\dd t^2)\]
with, as in the main text, $p(x,\delta)\dd x$ the unconditioned probability that there be an active segment in a $\dd x$-neighbourhood of $x$ lasting longer than $\delta$. Hence, if Markovianity holds, we have
\beqa
\lefteqn{\mathbb{E}[ T^\epsilon(x,t+\dd t)-T^\epsilon(x,t)|\mathcal{F}_t]} && \nonumber \\
&=&  -\epsilon\, \partial_x T^{\epsilon}(x,t)\, \dd t  + \eta \dd t\, \big( T^{-\epsilon}(x,t) - T^{\epsilon}(x)\big) +O(\dd t^2). \nonumber
\eeqa
Now, by taking the mean, and using the fundamental property of conditioned expectations, $\expec[ \expec [(\cdots)\vert \mathcal{F}_t]]=\expec [(\cdots)]$, we get:
\beqs
\partial_t \mathbb{E}[ T^\epsilon(x,t)] =  -\epsilon\, \partial_x \expec [T^{\epsilon}(x,t)]
+ \eta \, \expec [ T^{-\epsilon}(x) - T^{\epsilon}(x,t) ].
\eeqs  
This is equivalent to the hydrodynamic equations (\ref{eq:dissipA},\ref{eq:dissipB}).

 \bigskip
 
\underline{\it On the non-equilibrium one-point functions.}

By construction, the imbalanced initial conditions are such that the energy density has a jump at the origin, $h_{t=0}(x)=h_l$ on the left for $x<0$ and  $h_{t=0}(x)=h_r$ for $x>0$, but with vanishing momentum density, $p_{t=0}(x)=0$. Thanks to the local energy conservation, this translates into two initial conditions for the momentum density $p_{t=0}(x)=0$ and $\partial_t p_{t=0}(x)= (h_l-h_r)\, \delta(x)$. It is then easy to solve eq. (\ref{eq:p-seul}) by Fourier transform. The solution is $p(x,t)= (h_l-h_r)\, G(x,t)$ with $G(x,t)$ given in eq. (\ref{eq:def-G}).

One may notice that eq. (\ref{eq:p-seul}) can be mapped into the Klein-Gordon equation with negative square mass, $(\partial_t^2-\partial_x^2)q=q$, via the mapping $p(x,t)=e^{-\eta t}q(x,t)$.

The energy density $h(x,t)$ is found by integrating back the conservation law $\partial_t h(x,t)+\partial_x p(x,t)=0$ with initial condition $h_l$ for $x<0$ and $h_r$ for $x>0$. In particular its late time limit is $h_{t=0}(x)-\int_0^\infty \dd s\, \partial_xp(x,s)$. It is homogeneous, $x$-independent, and equal to $h_{eq}=\frac{1}{2}(h_r+h_l)$ as expected because $\int_0^\infty \dd s\, \partial_x G(x,s)=-\frac{1}{2}\,\mathrm{sign}(x)$.

Let us now prove the light-cone effect and the exponential decay. It is simpler to change variables, taking into account the relativistic character of the equations. For $0<k<\eta$, let $k=\eta\cos a$ with $a\in[0,\frac{\pi}{2}]$, for $k>\eta$, let $k=\eta\cosh a$ with $a\in[0,\infty]$. Hence, $G(x,t)=G_<(x,t)+G_>(x,t)$ with
\beqs
 G_<(x,t) &=& e^{-\eta t} \int_0^{\frac{\pi}{2}} \frac{\dd a}{\pi}\, \sinh(\eta t\sin a)\, \cos(\eta x\cos a),\\
G_>(x,t) &=& e^{-\eta t} \int_0^\infty \frac{\dd a}{\pi}\, \sin(\eta t\sinh a)\, \cos(\eta x\cosh a).
\eeqs
The light cone effect is proved by representing $G_>$ and $G_<$ as contour integrals. For $x>t>0$ outside the light-cone, we set $x=r\cosh u$, $t=r\sinh u$ with $u>0$, and rewrite $G_<$ and $G_>$ as :
\beqs
G_>(x,t) &=& e^{-\eta t}\, \Big( - \int_{-u}^{+u} + \int_{-u+\ii\pi}^{+u+i\pi}\Big) \frac{\dd a}{4\ii\pi}\, e^{\ii\eta r\cosh a}, \\
G_<(x,t) &=& e^{-\eta t}\, \Big( - \int_{+u}^{+u+\ii\pi} + \int_{-u}^{-u+i\pi}\Big) \frac{\dd a}{4\ii\pi}\, e^{\ii\eta r\cosh a} .
\eeqs
Adding the two  contributions gives a closed contour and hence the vanishing of the sum.

The asymptotic behaviour inside the light-cone is obtained from a saddle approximation. Let us consider the asymptotic along a ray $x=vt$ with $v=\tanh\gamma$. The ballistic component is negligible at large time. We have 
\[   \frac{G_>(vt,t)}{G_<(vt,t)} \simeq e^{-2\eta t/\cosh\gamma} \to 0\quad(t\to\infty) .\] 
The dominating contribution clearly comes from the diffusive component $G_< (vt,t)$, which takes the form
\beqs G_<(vt,t) &=& e^{-\eta t}\,  \int_{-\pi/2}^{\pi/2} \frac{\dd a}{2\pi}\, \big[ e^{ \eta t\frac{\sin (a+\ii\gamma)}{\cosh\gamma}} + e^{ \eta t\frac{\sin (a-\ii\gamma)}{\cosh\gamma}} \big].
\eeqs
After contour deformation $a\mapsto a-\ii\gamma$ and $a\mapsto a+\ii\gamma$ for the first and second term inside the integral, respectively, the saddle point is at $a=\pi/2$. Hence, inside the light cone ($v=\tanh\gamma\in(-1,1)$)
\[ G(vt,t) \simeq G_<(vt,t) \simeq  \mathrm{const.}\, \frac{e^{-\eta t}e^{+\eta t/\cosh\gamma}}{\sqrt{t/\cosh\gamma}} \quad(t\to\infty).\]

Finally, let us define the normalized momentum at each time by $\hat p(v,t):= {\h P_t}^{-1}\, G(vt,t)$, with $\h P_t$ the current mean spatial average $\h P_t=t^{-1}\int_{-t}^t \dd x\, G(x,t)$, so that $\int_{-1}^{+1} \dd v\, \hat p(v,t)=1$ for all time $t$. It is easy to verify that at large time this concentrate to a delta function: $\lim_{t\to \infty} \hat p(v,t)= \delta(v)$. This is yet another signature of localisation: on the ballistic scale of order $t$ the momentum flow remains concentrated at the origin.
\bigskip

{\underline{\it Spatial fluct\\uations of momentum}}

At large times, spatial averages reproduce stochastic averages. Indeed, in any realization, a backward random trajectory from a point $x$ gets to a position $x_0$ that is not correlated with $x$. This leads to $\int_{a}^b \dd x\,\bra p(x)\ket_{t\to\infty} = 0$. It also indicates that the integration of the square of the quantum-average momentum is nonzero, and more precisely given by
\beq
	\lim_{t\to\infty} \frc{\int_{a}^b \dd x\,\big(\bra p(x)\ket_{t}\big)^2}{b-a} =
	\frc1{16}\sum_{\ep,\ep',\eta,\eta'} \lt(
	\bra T^\ep\ket_\eta - \bra T^{\ep'}\ket_{\eta'}\rt)^2.
\eeq
This is because $\bra p(x)\ket_t$ may take, at large $t$, any one of the 16 possible values $\bra T^\ep\ket_\eta - \bra T^{\ep'}\ket_{\eta'}$ with equal probability.

\vfill


\begin{thebibliography}{99}

\bibitem{Anderson} P.W. Anderson, Phys. Rev. 109, 1492 (1958).

\bibitem{Localisation} F. Evers, A.D. Mirlin, Rev. Mod. Phys. 80, 1355 (2008).

\bibitem{Many-Body} A. Pal, D.A. Huse, Phys. Rev. B 82, 174411 (2010);\\
	R. Nandkishore, D.A. Huse, Ann. Rev. Cond. Mat. Phys. 6,15-38 (2015);\\
E. Altman, R. Vosk, Ann. Rev.Cond. Matt. Phys. 6, 383-409 (2015);\\
 R. Vasseur, J. E. Moore, J. Stat. Mech. 2016, 064010 (2016).

\bibitem{CFT} Ph. Di Francesco, P. Mathieu, D. Senechal, {\it Conformal Field Theory}, Springer-NY, (1997).

\bibitem{OQBM} 
M. Bauer, D. Bernard, A. Tilloy, J. Stat. Mech. 2014, P09001 (2014). 

\bibitem{Breuer} H.P. Breuer and F. Petruccione, {\it The Theory of Open Quantum Systems}, Oxford Univ. Press, 2006.

\bibitem{diff0} M. Ljubotina, M. Znidaric, T. Prosen, Nat. Commun. 8, 16117 (2017).

\bibitem{diff}  N. Bloembergen, Physica 15,  386 (1949);\\ P.G. de Gennes, J. Phys. Chem. Solids, 4, 223 (1958);\\ K. Fabricius, B. M. McCoy, Phys. Rev. B 57, 8340 (1998);\\ J. Sirker, R. G. Pereira, I. Affleck, Phys. Rev. B 83, 035115 (2011).

\bibitem{CP-evol} E.B. Davies, {\it Quantum Theory of Open Systems}, Academic Press, 1976.


\bibitem{Lindblad} G. Lindblad, Commun. Math. Phys. 48, 119 (1976).

\bibitem{Cattaneo} C. Cattaneo, Atti del Seminario matematico e fisico della Universit\`a di Modena 3, 3-21 (1948); C. R. Acad. Sci. Paris 247, 431-433  (1958).

\bibitem{DF} This is $D_F=v_F^2/\eta$ with a Fermi velocity $v_F\neq1$.

\bibitem{BD-all} D. Bernard, B. Doyon, J. Phys. A 45, 362001 (2012);\\
D. Bernard, B. Doyon, J. Stat. Mech. 064005 (2016).

\bibitem{BD-hydro} D. Bernard, B. Doyon, J. Stat. Mech. 033104 (2016).

\bibitem{GMMP} I.V. Gornyi, A.D. Mirlin, M. M\"uller, D.G. Polyakov, Ann. Phys.  529, 1600365 (2017).

\end{thebibliography}
\end{document}